\documentclass[showpacs,preprintnumbers,amsmath,amssymb,twocolumn]{revtex4}
\usepackage{amsfonts}

\usepackage{graphicx}
\usepackage{dcolumn}
\usepackage{bm}
\usepackage{amssymb}
\usepackage{array}
\usepackage{longtable}

\begin{document}

\preprint{APS/123-QED}

\title{Determining mean first-passage time on a class of treelike regular fractals}

\author{Yuan Lin}
\author{Bin Wu}
\author{Zhongzhi Zhang}
\email{zhangzz@fudan.edu.cn}
\homepage{http://homepage.fudan.edu.cn/~zhangzz/}

\affiliation {School of Computer Science, Fudan University, Shanghai
200433, China}

\affiliation {Shanghai Key Lab of Intelligent Information
Processing, Fudan University, Shanghai 200433, China}




\date{\today}

\begin{abstract}
Relatively general techniques for computing mean first-passage time
(MFPT) of random walks on networks with a specific property are very
useful, since a universal method for calculating MFPT on general
graphs is not available because of their complexity and diversity.
In this paper, we present techniques for explicitly determining the
partial mean first-passage time (PMFPT), i.e., the average of MFPTs
to a given target averaged over all possible starting positions, and
the entire mean first-passage time (EMFPT), which is the average of
MFPTs over all pairs of nodes on regular treelike fractals. We
describe the processes with a family of regular fractals with
treelike structure. The proposed fractals include the $T$ fractal
and the Peano basin fractal as their special cases. We provide a
formula for MFPT between two directly connected nodes in general
trees on the basis of which we derive an exact expression for PMFPT
to the central node in the fractals. Moreover, we give a technique
for calculating EMFPT, which is based on the relationship between
characteristic polynomials of the fractals at different generations
and avoids the computation of eigenvalues of the characteristic
polynomials. Making use of the proposed methods, we obtain
analytically the closed-form solutions to PMFPT and EMFPT on the
fractals and show how they scale with the number of nodes. In
addition, to exhibit the generality of our methods, we also apply
them to the Vicsek fractals and the iterative scale-free fractal
tree and recover the results previously obtained.


\end{abstract}

\pacs{05.40.Fb, 61.43.Hv, 89.75.Hc, 05.60.Cd}
\maketitle


\section{Introduction}

As an integral ingredient of nonlinear science, fractals have
attracted an increasing attention in physics and other scientific
fields~\cite{HaBe87,BeHa00} not only because of the striking beauty
intrinsic in their structure~\cite{Ma82} but also due to the
significant impact of the idea of fractals on a large variety of
scientific disciplines, such as astrophysics, plasma physics,
optics, economy, ecology, and so on~\cite{AgViSa09}. Among various
fractal categories, regular fractals constitute an important family
of fractals. Frequently cited examples include the Cantor
set~\cite{Ca1881}, the Koch curve~\cite{Ko1906}, the Sierpinski
gasket~\cite{Si1915}, the Vicsek fractals~\cite{Vi83}, the $T$
fractal~\cite{KaRe86}, etc. These structures have been a focus of
research object~\cite{Fa03}. One of the main justifications for
studying regular fractals is that many problems can be exactly
solvable on regular fractals~\cite{ScScGi97}, the explicit solutions
for which provide useful insight different from that of approximate
solutions for random fractals.

One of the ultimate goals of research efforts on fractals is to
unveil how their underlying geometrical and structural features
affect dynamical processes and critical phenomenon on
them~\cite{DoGoMe08}. Among a variety of dynamics, due to their
simplicity and wide range of applications~\cite{We1994}, random
walks play a central role in many branches of sciences and
engineering and have gained a considerable attention within the
scientific community~\cite{MeKl00,BuCa05}. A key quantity related to
random walks is mean first-passage time (MFPT)~\cite{Re01,NoRi04}
that is the expected time for the walker starting off from a source
node to first reach a given target node. In addition to its role as
a basic quantitative measure of the transportation efficiency, MFPT
also encodes useful information of other quantities concerned with
random walks~\cite{Re01}.

Concerted efforts have been devoted to study MFPT in fractals in
order to obtain the scaling of MFPT with system
size~\cite{CoBeTeVoKl07}. For instance, MFPT associated to a given
starting node has been discussed by many
authors~\cite{CoBeMo05,CoBeMo07} since it depends on the starting
position and is an important issue in its own right. On the other
hand, recent papers have addressed the explicit determination of
partial mean first-passage time (PMFPT) on some regular fractals,
defined as the average of MFPTs to a selected target node averaged
over all possible starting nodes, such as the Sierpinski
gasket~\cite{KaBa02PRE,KaBa02IJBC}, the $T$
fractal~\cite{Ag08,HaRo08}, the iterative scale-free treelike
fractal~\cite{ZhXiZhGaGu09}, as well as the hierarchical lattice
fractals~\cite{ZhXiZhLiGu09,TeBeVo09}. Besides, entire mean
first-passage time (EMFPT), i.e., the average of MFPTs over all
pairs of nodes, has been computed for the $T$
fractal~\cite{ZhLiZhWuGu09}, the Vicsek
fractals~\cite{ZhWjZhZhGuWa10}, and some other regular
fractals~\cite{CoBeTeVoKl07}. Thus far, most techniques used to
calculate PMFPT or EMFPT are only applicable to specific structure;
universal (even relatively general) methods for computing MFPTs are
much less~\cite{NoRi04,CoMi10}.

In this paper, we develop techniques for the explicit determination
of PMFPT and EMFPT on regular treelike fractals, which are also
expected to be valid to other deterministic media. For PMFPT, we
provide a universal formula for MFPT between two adjacent nodes in
generic trees based on which one can determine PMFPT between a given
node and all other nodes. For EMFPT, by considering only the
relationship between characteristic polynomials at different
iterations, we can find exactly the analytical expression for EMFPT,
which needs only local information of characteristic polynomials and
can avoid the laborious explicit computation of the eigenvalues.

To illustrate the calculation processes, we put forward a family of
treelike regular fractals, which include the famous $T$ fractal and
the Peano basin fractal~\cite{BaDeVe09} as their particular cases.
Using the presented methods, we first study the random-walk problem
with a trap fixed at the central node of the fractals and obtain the
PMFPT to the trap; then, we address the case that the trap is
uniformly selected among all nodes and derive the rigorous solution
to EMFPT between all node pairs. From the obtained analytical
results, we give the scalings for both PMFPT and EMFPT and show that
they both increase as a power-law function of the number of nodes,
with the exponent larger than 1 but smaller than 2. 

\begin{figure}
\begin{center}
\includegraphics[width=.90\linewidth,trim=0 0 0 0]{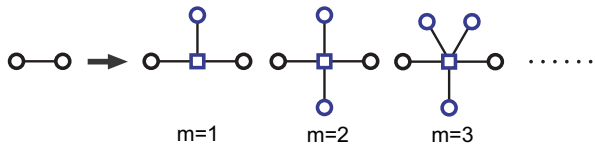}
\caption{ (Color online) Iterative construction method of the
treelike regular fractals. The next generation is obtained by
performing the operation shown to the right of the arrow.}
\label{cons}
\end{center}
\end{figure}

\section{Constructions and properties of the treelike fractals \label{model}}

\begin{figure}
\begin{center}
\includegraphics[width=0.9\linewidth,trim=0 20 30 0]{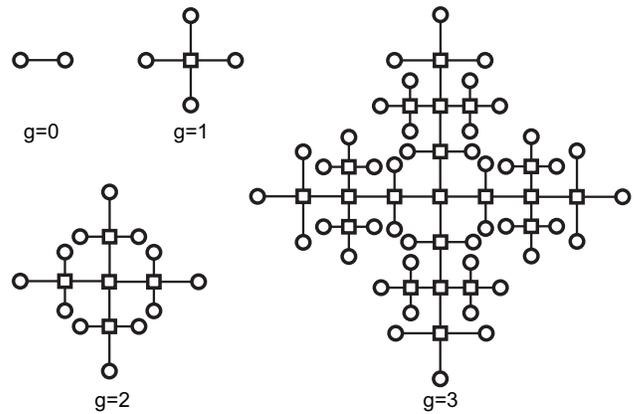}
\end{center}
\caption[kurzform]{Illustration for the growing process for a
special fractal corresponding to the case of $m=2$. A circle
expresses an external node, while a square denotes an internal
node.} \label{network}
\end{figure}

Let us first introduce the model for the treelike fractals that are
built in an iterative way and controlled by a positive-integer
parameter $m$ (i.e., $m \ge 1$). We denote by $T_{g}$ ($g\ge0$) the
treelike fractals after $g$ iterations. Notice that the number of
iterations also represents the generation of the treelike fractals.
Initially ($g=0$), $T_{0}$ is an edge connecting two nodes. For $g
\ge 1$, $T_{g}$ is obtained from $T_{g-1}$ by performing the
following operations on every edge in $T_{g-1}$ as shown in
Fig.~\ref{cons}: replace the edge by a path of two links long, with
the two end points of the path being the same end points of the
original edge, then attach $m$ new nodes to the middle node of the
path. Figure~\ref{network} illustrates the construction process of a
particular fractal for the case of $m=2$, showing the first several
generations. Notice that there are several limiting cases of our
model. When $m = 1$, it is reduced to the $T$ fractal~\cite{KaRe86}.
When $m = 2$, it turns out to be the Peano basin
fractal~\cite{BaDeVe09}.

According to the construction algorithm, at each step the number of
edges in the treelike fractals increases by a factor of $m+2$. Thus,
we can easily know that the total number of edges of $T_{g}$ is
$E_{g}=(m+2)^{g}$ and that the total number of nodes (often named
network order) in $T_{g}$ is $N_{g}=E_{g}+1=(m+2)^{g}+1$. One can
partition all the $N_{g}$ nodes into two classes: nodes having
degree $m+2$ are called internal nodes and nodes with degree of 1
are named external nodes. Since at each generation, a preexisting
edge creates $m+1$ nodes: one of which has a degree of $m+2$, the
rest $m$ nodes have a degree of 1. Thus, at generation $g_i$ ($g_i
\geq 1$), the numbers of internal and external nodes generated at
this generation are $(m+2)^{g_i-1}$ and $m(m+2)^{g_i-1}$,
respectively.

On the other hand, it is obvious that after each iteration, the
diameter (namely, the maximum of shortest distances between all
pairs of nodes) of the fractals doubles. Thus, we have that the
fractal dimension of the treelike fractals is $d_{f}=\ln
(m+2)/\ln2$. In addition, as we will show below, for any two nodes
$i$ and $j$ at current generation, the MFPT from node $i$ to node
$j$ increases by a factor of $2(m+2)$ at next generation. Hence, the
random-walk dimension of the fractals is
$d_{w}=\ln[2(m+2)]/\ln2=1+d_{f}$, and their spectral dimension is
$\widetilde{d}=2d_{f}/d_{w}=2\ln
(m+2)/\ln[2(m+2)]=2d_{f}/(1+d_{f})$.

\begin{figure}
\begin{center}
\includegraphics[width=0.6\linewidth,trim=0 20 0 0]{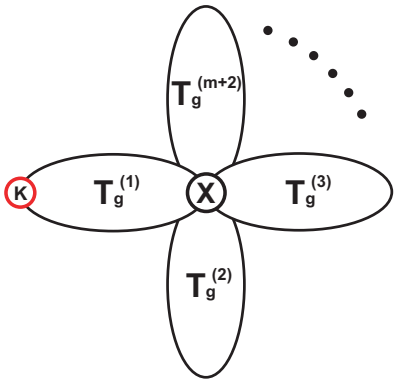}
\end{center}
\caption[kurzform]{(Color online) An alternative construction method
of the treelike fractals. The fractals after $g+1$ iterations,
$T_{g+1}$, can be obtained by merging $m+2$ copies of $T_{g}$,
sequentially denoted by $T_{g}^{(1)}$, $T_{g}^{(2)}$, $\ldots$,
$T_{g}^{(m+1)}$, and $T_{g}^{(m+2)}$, where $X$ is the innermost
node, while $K$ is an outermost node.}\label{Const2}
\end{figure}

It is worth mentioning that the treelike fractals can also be
alternatively constructed using another method. If we define the
central node (see Fig.~\ref{network} for an example) as the
innermost node and the nodes farthest from the central node as the
outermost nodes, then the second generating algorithm for the
fractals can be described as follows: Given the generation $g$,
$T_{g+1}$ can be obtained by joining $m+2$ replicas of $T_{g}$ (see
Fig.~\ref{Const2}). That is to say, to obtain $T_{g+1}$ one can
merge together the discrete outermost nodes of the $m+2$ copies of
$T_{g}$. The $m+2$ outermost nodes belonging to separate copies
merge into a single new node, which is then the innermost node in
$T_{g+1}$.

After introducing the construction algorithms and the properties of
the treelike fractals, in what follows we will study the random-walk
dynamics on the fractal family $T_{g}$. First, we will investigate
random walks with a single immobile trap located on the innermost
node. Then, we will continue to address random walks with the trap
selected randomly (i.e., distributed uniformly) among all nodes.

\section{random walks with a trap fixed on the central node}

In this section we focus on a particular case of random
walks---trapping problem---on the family of treelike fractals
$T_{g}$ with the trap or perfect absorber placed on the central node
of the regular fractals. In the process of the random walks, at each
time step, the walker, starting from its current location, jumps to
any of its neighbor nodes with identical probability. For the
convenience of description, we label by 1 the central node of
$T_{g}$, while all other nodes are labeled consecutively as 2, 3,
$\ldots$, $N_{g}-1$, and $N_{g}$. Let $F_{i}(g)$ be the MFPT, also
called trapping time, for a walker, staring from node $i$ to first
arrive at the trap. What we are concerned is the PMFPT presented as
$\langle F \rangle_{g}$, which is defined as the average of
$F_{i}(g)$ over all starting nodes distributed uniformly over nodes
in the proposed fractals other than the trap. PMFPT is also often
called mean trapping time or mean time to absorption~\cite{Ag08}. By
definition, $\langle F \rangle_{g}$ is given by
\begin{equation}\label{MFPT}
\langle F \rangle_{g}=\frac{1}{N_g-1}\sum_{i=2}^{N_g} F_i(g)\,.
\end{equation}
The main purpose of this section is to determine explicitly $\langle
F \rangle_{g}$ and show how $\langle F \rangle_{g}$ scales with
network order.

\subsection{Mean first-passage time between two adjacent nodes in general trees}

Here, we provide a universal formula for MFPT between two nodes
directly connected by an edge in a treelike network, which is very
helpful for the following derivation of $\langle F \rangle_{g}$. Let
us consider a general connected tree. We denote by $e=(u,v)$ an edge
in the tree connecting two nodes $u$ and $v$. Obviously, if we
remove the edge $(u,v)$, the tree will be divided into two small
subtrees: one subtree includes node $u$ and the other contains node
$v$. We use $C_{u < v}$ to denote the number of nodes in the subtree
containing node $u$. Actually, $C_{u < v}$ is the number of nodes in
the original tree lying closer to $u$ than to $v$, including $u$
itself. Furthermore, let $F_{uv}$ denote the MFPT from $u$ to $v$.
Then, $F_{uv}$ can be written in terms of $C_{u < v}$ as
\begin{equation}\label{A1}
F_{uv}=2C_{u < v}-1.
\end{equation}

Equation~(\ref{A1}) can be easily proved inductively. If $u$ is a
leaf node (namely, node with degree 1), it is obvious that $C_{u <
v}=1$ and $F_{uv}=1$, and thus Eq.~(\ref{A1}) holds. We proceed to
consider the case that $u$ is a non-leaf node. For this case, to
prove the validity of Eq.~(\ref{A1}), we consider the tree as a
rooted one with node $v$ being its root, then $v$ is the father of
$u$ and $C_{u < v}$ is in fact the number of nodes in the subtree
whose root is $u$. Suppose that for a non-leaf node $u$,
Eq.~(\ref{A1}) is true for all its children, the set of which is
denoted by $\Omega_u$. In other words, for an arbitrary node $x$ in
$\Omega_u$, the following relation holds:
\begin{equation}\label{A2}
F_{xu}=2C_{x < u}-1\,.
\end{equation}
Then, the MFPT $F_{uv}$ from node $u$ to its father node $v$ can be
calculated by
\begin{equation}\label{A3}
F_{uv}=\frac{1}{\delta}+\frac{1}{\delta}\sum_{x \in \Omega_u}
(1+F_{xv})\,,
\end{equation}
where $\delta$ is the degree of node $u$.

The first term on the right-hand side (rhs) of Eq.~(\ref{A3})
explains the case that the walker, starting off from node $u$, jumps
directly to node $v$ in one single step with probability
$\frac{1}{\delta}$. The second term accounts for another case that
the walker first reaches a child node $x$ of node $u$ in one step,
and then it takes more $F_{xv}$ steps to first hit the target node
$v$. Equation~(\ref{A3}) can be rewritten as
\begin{equation}\label{A4}
F_{uv}=\frac{1}{\delta}+\frac{1}{\delta}\sum_{x \in \Omega_u}
(1+F_{xu}+F_{uv})\,,
\end{equation}
which leads to
\begin{eqnarray}\label{A5}
F_{uv} &=&1+\sum_{x \in \Omega_u}
(1+F_{xu})=\delta+\sum_{x \in \Omega_u}(2C_{x < u}-1)\nonumber\\
&=&2C_{u <v}-1,
\end{eqnarray}
where the assumption, viz. Eq.~(\ref{A2}), was used. Thus, we have
proved Eq.~(\ref{A1}), which is a basic characteristic for random
walks on a tree and is useful for the following computation of the
key quantity $\langle F \rangle_{g}$.

\subsection{Evolution law for mean first-passage time in the fractals}


Let $F_{ij}(g)$ be the MFPT of random walks on $T_{g}$, staring from
node $i$, to first reach node $j$. Let $(u,v)$ be an edge connecting
two nodes $u$ and $v$ in $T_{g}$. Next, we will give a relation
between $F_{uv}(g+1)$ and $F_{uv}(g)$. To this end, we look upon the
treelike regular fractals $T_{g}$ as rooted trees with node $v$ as
the root, and thus $v$ is the father of $u$. Note that in the
evolution of the fractals, we suppose that node $v$ is always the
root. In addition, for $T_{g}$ we use $C_{u}(g)$ and $E_{u}(g)$ to
represent the numbers of nodes and edges in the subtree, whose root
is node $u$. Clearly,
\begin{equation}\label{B1}
E_{u} (g)=C_{u}(g)-1\,.
\end{equation}
On the other hand, by construction of the fractals, each edge in
$T_{g}$ will be replaced by $m+2$ new edges at iteration $g+1$.
Then, we can easily have
\begin{equation}\label{B2}
E_{u}(g+1)=(m+2)E_{u}(g)\,,
\end{equation}
from which we can obtain the recursion relation between $C_{u}(g+1)$
and $C_{u}(g)$. By definition,
\begin{eqnarray}\label{B3}
C_{u}(g+1)&=&E_{u}(g+1)+1=(m+2)E_{u}(g)+1\nonumber \\
&=&(m+2)C_{u}(g)-m-1\,.
\end{eqnarray}

\begin{figure}
\begin{center}
\includegraphics[width=0.6\linewidth,trim=0 20 0 0]{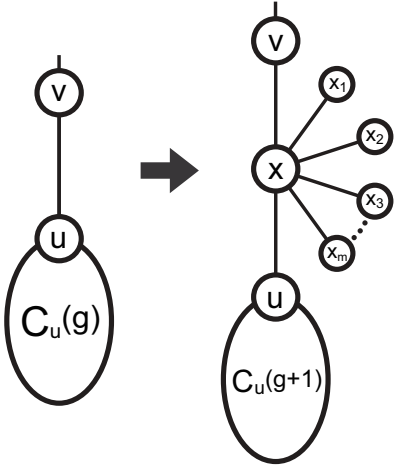}
\end{center}
\caption[kurzform]{Illustration for the evolution of the
fractals.}\label{two}
\end{figure}

Now we begin to derive the relation governing $F_{uv}(g+1)$ and
$F_{uv}(g)$. According to the general result given in
Eq.~(\ref{A1}), we have
\begin{equation}\label{B4}
F_{uv}(g)=2C_{u}(g)-1\,.
\end{equation}
Notice that at generation $g+1$, $v$ is no longer the father of $u$
but still an ancestor of $u$. According to the construction
algorithm of the fractals, the edge in $T_{g}$ connecting nodes $u$
and $v$ will generate $m+1$ new nodes at generation $g+1$ (see
Fig.~\ref{two}): one node $x$ has a degree $m+2$, while the other
$m$ nodes denoted by $x_i$ $(i=1,2,\ldots, m)$ have the same degree
1. After this iteration, the father of $u$ becomes $x$ and $x$ is a
child of $v$. Then, for a random walker in $T_{g+1}$, if it wants to
transfer from $u$ to $v$, it must pass by node $x$. Therefore,
\begin{eqnarray}\label{B5}
F_{uv}(g+1)&=&F_{ux}(g+1)+F_{xv}(g+1)\nonumber \\
&=&2C_{u}(g+1)+2C_{x}(g+1)-2.
\end{eqnarray}
Then, the determination of $F_{uv}(g+1)$ is reduced to finding
$C_{u}(g+1)$ and $C_{x}(g+1)$.

Since $C_{u}(g+1)$ has been determined above, we only need to find
$C_{x}(g+1)$. According to the structure of the fractals, it is
obvious that
\begin{equation}\label{B6}
C_{x}(g+1)=C_{u}(g+1)+m+1=(m+2)C_{u}(g)\,,
\end{equation}
where the term $m+1$ between the two equal marks describes the
number of nodes $x$ and $x_i$ $(i=1,2,\ldots, m)$. Plugging
Eqs.~(\ref{B3}) and~(\ref{B6}) into Eq.~(\ref{B5}), we obtain
\begin{eqnarray}\label{B7}
F_{uv}(g+1)&=&4(m+2)C_{u}(g)-2m-4\nonumber\\
&=&2\left(m+2\right)F_{uv}(g)\,,
\end{eqnarray}
in which Eq.~(\ref{B4}) has been made use of.

Equation~(\ref{B7}) tells us that for any two adjacent nodes $u$ and
$v$ in the fractals at a given generation, the MFPT from one of them
to the other will increase by a factor of $2(m+2)$ at the next
generation. Since the considered fractals have a treelike structure,
for any pair of two nodes $i$ and $j$, adjacent or not, the MFPT
between them obeys the relation:
\begin{equation}\label{B8}
F_{ij}(g+1)=2(m+2)F_{ij}(g)\,,
\end{equation}
which is a basic property of random walks in the regular treelike
fractals, dominating the evolution of MFPT between any couple of
nodes. The relation described by Eq.~(\ref{B8}) is very important
using which we will derive the rigorous formula for $\langle F
\rangle_{g}$.

\subsection{Exact solution to partial mean first-passage time in $T_{g}$}

Having obtained the evolution rule of MFPT for random walks in the
fractals, we now determine the mean time to absorption averaged over
all non-trap nodes in $T_{g}$. To attain this goal, we represent the
set of nodes in $T_{g}$ as $\Lambda_{g}$, and denote the set of
those nodes created at generation $g$ by $\bar{\Lambda}_{g}$.
Obviously, $\Lambda_{g}=\Lambda_{g-1}+\bar{\Lambda}_{g}$. For the
convenience of description, we define the following two quantities
for $n \leq g$:
\begin{equation}\label{C1}
F_{n}^{\rm tot}(g)=\sum_{i\in \Lambda_{n}}F_{i}(g)\,,
\end{equation}
and
\begin{equation}\label{C2}
\bar{F}_{n}^{\rm tot}(g)=\sum_{i\in \bar{\Lambda}_{n}}F_{i}(g)\,.
\end{equation}
Thus, we have
\begin{equation}\label{C3}
F_{g}^{\rm tot}(g)=2(m+2)F_{g-1}^{\rm tot}(g)+\bar{F}_{g}^{\rm
tot}(g)\,,
\end{equation}
where we have used Eq.~(\ref{B8}). Thus, to obtain $F_{g}^{\rm
tot}(g)$, we should first determine the quantity $\bar{F}_{g}^{\rm
tot}(g)$.

To find $\bar{F}_{g}^{\rm tot}(g)$, we separate set
$\bar{\Lambda}_{g}$ of the nodes created at generation $g$ into two
subsets $\bar{\Lambda}_{g}^{\rm int}$ and $\bar{\Lambda}_{g}^{\rm
ext}$, such that $\bar{\Lambda}_{g}=\bar{\Lambda}_{g}^{\rm
ext}\cup\bar{\Lambda}_{g}^{\rm int}$, where $\bar{\Lambda}_{g}^{\rm
int}$ is the set of internal nodes and $\bar{\Lambda}_{g}^{\rm ext}$
is the set of external nodes. As shown in Sec. \ref{model}, the
cardinalities (the cardinality of a set is the number of nodes in
the set) of the two subsets, denoted  by $|\bar{\Lambda}_{g}^{\rm
ext}|$ and $|\bar{\Lambda}_{g}^{\rm int}|$, are
$|\bar{\Lambda}_{g}^{\rm int}|=(m+2)^{g-1}$ and
$|\bar{\Lambda}_{g}^{\rm ext}|=m(m+2)^{g-1}$, respectively. Clearly,
the two variables satisfy the following relation:
\begin{equation}\label{C4}
|\bar{\Lambda}_{g}^{\rm ext}|=m|\bar{\Lambda}_{g}^{\rm int}|.
\end{equation}

Then, the quantity $\bar{F}_{g}^{\rm tot}(g)$ can be rewritten as
\begin{equation}\label{C5}
\bar{F}_{g}^{\rm tot}(g)=\sum_{i\in \bar{\Lambda}_{g}^{\rm
int}}F_{i}(g)+\sum_{i\in \bar{\Lambda}_{g}^{\rm ext}}F_{i}(g)\,,
\end{equation}
which shows that to determine $\bar{F}_{g}^{\rm tot}(g)$, one may
alternatively find the two quantities on the rhs of the equal mark.
We begin from determining the first sum term $\sum_{i\in
\bar{\Lambda}_{g}^{\rm int}}F_{i}(g)$.

Since at a given generation $g_i$  ($g_i \geq 2$), any internal node
has a degree of $m+2$ and two internal nodes in
$\bar{\Lambda}_{g_i}^{\rm int}$ are not directly connected to each
other, we have that any of the $E_{g_i}$ edges in network $T_{g_i}$
must have one (and only one) end point belonging to
$\bar{\Lambda}_{g_i}^{\rm int}$. In addition, by construction any
internal node in $\bar{\Lambda}_{g_i+1}^{\rm int}$ is linked to both
ends of the edge preexisting at generation $g_i$ that generated the
internal node, which means that any internal node in
$\bar{\Lambda}_{g_i+1}^{\rm int}$ must have one (and only one)
neighbor belonging to $\bar{\Lambda}_{g_i}^{\rm int}$. Furthermore,
we can know that at generation $g_i+1$, the $m+2$ neighbors of any
internal node in $\bar{\Lambda}_{g_i}^{\rm int}$ are all internal
nodes in $\bar{\Lambda}_{g_i+1}^{\rm int}$, and for any two internal
nodes in $\bar{\Lambda}_{g_i}^{\rm int}$, their neighbors are
completely different. From this information, we can obtain a
recursion relation for $\sum_{i\in \bar{\Lambda}_{g}^{\rm
int}}F_{i}(g)$.

Consider the random walks on the fractals $T_g$. For an internal
node $i_{g-1}^{\rm int}$ in $\bar{\Lambda}_{g-1}^{\rm int}$ and all
its $m+2$ neighbors $i_{g,x}^{\rm int}$ ($x=1,2,\ldots, m+2$)
belonging to $\bar{\Lambda}_{g}^{\rm int}$, their trapping times
obey the following relation:
\begin{equation}\label{C6}
F_{i_{g-1}^{\rm int}}(g)=1+\frac{1}{m+2}\sum_{x=1}^{m+2}
F_{i_{g,x}^{\rm int}}(g)\,.
\end{equation}
Summing Eq.~(\ref{C6}) over all internal nodes in
$\bar{\Lambda}_{g-1}^{\rm int}$, we obtain
\begin{equation}\label{C7}
\sum_{i\in\bar{\Lambda}_{g-1}^{\rm
int}}F_{i}(g)=|\bar{\Lambda}_{g-1}^{\rm
int}|+\frac{1}{m+2}\sum_{i\in\bar{\Lambda}_{g}^{\rm int}}F_{i}(g)\,,
\end{equation}
which may be rewritten as
\begin{eqnarray}\label{C8}
\sum_{i\in\bar{\Lambda}_{g}^{\rm
int}}F_{i}(g)&=&(m+2)\sum_{i\in\bar{\Lambda}_{g-1}^{\rm
int}}F_{i}(g)-(m+2)|\bar{\Lambda}_{g-1}^{\rm int}|\nonumber\\
&=&2\left(m+2\right)^{2}\sum_{i\in\bar{\Lambda}_{g-1}^{\rm
int}}F_{i}(g-1)-(m+2)^{g-1}.\nonumber\\
\end{eqnarray}
Using $\sum_{i\in\bar{\Lambda}_{2}^{\rm int}}F_{i}(2)=(2m+3)(m+2)$,
Eq.~(\ref{C8}) is resolved by induction,
\begin{equation}\label{C9}
\sum_{i\in\bar{\Lambda}_{g}^{\rm
int}}F_{i}(g)=\frac{(m+2)^{g-2}}{3+2m} [2+m+2^{g}(m+1)(m+2)^{g}]\,.
\end{equation}

After obtaining the first sum $\sum_{i\in \bar{\Lambda}_{g}^{\rm
int}}F_{i}(g)$ in Eq.~(\ref{C5}), we proceed to determine the second
sum $\sum_{i\in \bar{\Lambda}_{g}^{\rm ext}}F_{i}(g)$. A random
walker staring from a node $i_{\rm ext}$ belonging to
$\bar{\Lambda}_{g}^{\rm ext}$ must be on its only neighbor $i_{\rm
int}$---an internal node in $\sum_{i\in \bar{\Lambda}_{g}^{\rm
int}}F_{i}(g)$. Thus,
\begin{equation}\label{C10}
F_{i_{\rm ext}}(g)=F_{i_{\rm int}}(g)+1.
\end{equation}
Summing Eq.~(\ref{C10}) over all external nodes in
$\bar{\Lambda}_{g}^{\rm ext}$ gives rise to
\begin{equation}\label{C11}
\sum_{i\in \bar{\Lambda}_{g}^{\rm ext}}F_{i}(g)=m\, \sum_{i\in
\bar{\Lambda}_{g}^{\rm int}}F_{i}(g)+|\bar{\Lambda}^{\rm
ext}_{g}|\,,
\end{equation}
where we have considered Eq.~(\ref{C4}). Substituting Eq.~(\ref{C9})
for $\sum_{i\in\bar{\Lambda}_{g}^{\rm int}}F_{i}(g)$ and
$|\bar{\Lambda}^{\rm ext}_{g}|=m(m+2^{g-1})$ into Eq.~(\ref{C11}),
we have
\begin{eqnarray}\label{C12}
\sum_{i\in\bar{\Lambda}_{g}^{\rm ext}}F_{i}(g)&=&\frac{m\left(2+m\right)^{g-2}}{3+2m}\left(2+m+2^{g}\left(m+1\right)\left(2+m\right)^{g}\right)\nonumber\\
&\quad&+m\left(m+2\right)^{g-1}\,.
\end{eqnarray}

With the results obtained for $\sum_{i\in \bar{\Lambda}_{g}^{\rm
int}}F_{i}(g)$ and $\sum_{i\in \bar{\Lambda}_{g}^{\rm
ext}}F_{i}(g)$, we can determine $F_{g}^{\rm tot}(g)$. Substituting
the two sums given in Eqs.~(\ref{C9}) and~(\ref{C12}) for
$\bar{F}_{g}^{\rm tot}(g)$ into Eq.~(\ref{C3}) yields to
\begin{eqnarray}\label{C13}
F_{g}^{\rm tot}(g)&=&2(m+2)F_{g-1}^{\rm tot}(g)+m(m+2)^{g-1}+\nonumber\\
&\quad&\frac{(m+1)(2+m)^{g-2}}{3+2m}\left[2+m+2^{g}(m+1)(2+m)^{g}\right].\nonumber\\
\end{eqnarray}
Considering the initial condition $F_{1}^{\rm tot}(1)=m+2$,
Eq.~(\ref{C13}) is resolved by induction to yield
\begin{eqnarray}\label{C14}
F_{g}^{\rm tot}(g)&=&\frac{(2+m)^{g}}{2(2m^{2}+7m+6)}\big[2^{g+1}(m+1)(2+m)^{g}\nonumber\\
&\quad&+(2m^{2}+5m+3)2^{g}-4m^{2}-8m-2\big].
\end{eqnarray}
Plugging the last expression into Eq.~(\ref{MFPT}), we arrive at the
explicit formula for the PMFPT on the studied treelike regular
fractals,
\begin{eqnarray}\label{C15}
\langle F\rangle_{g}&=&\frac{1}{N_{g}-1}\sum_{i=2}^{N_{g}}F_{i}(g)
=\frac{1}{N_{g}-1}F_{g}^{\rm tot}(g)\nonumber\\
&=&\frac{1}{2(2m^{2}+7m+6)}\big[2^{g+1}(m+1)(2+m)^{g}\nonumber\\
&\quad&+(2m^{2}+5m+3)2^{g}-4m^{2}-8m-2\big]\,.
\end{eqnarray}
When $m=1$, Eq.~(\ref{C15}) recovers the result previously obtained
in Refs.~\cite{Ag08} and \cite{HaRo08}, confirming that the solution
given by Eq.~(\ref{C15}) is right.



We continue to show how to represent PMFPT as a function of network
order $N_{g}$, with the aim to obtain the scaling between these two
quantities. Recalling $N_{g}=(m+2)^{g}+1$, we have
$g=\log_{m+2}(N_{g}-1)$. Hence, Eq.~(\ref{C15}) can be recast as
\begin{eqnarray}\label{C16}
\langle F \rangle_{g}&=&\frac{1}{2(2m^{2}+7m+6)}\big[2(m+1)(N_{g}-1)^{1+\ln2/\ln(m+2)}\nonumber\\
&\quad&+(2m^{2}+5m+3)(N_{g}-1)^{\ln2/\ln(m+2)}\nonumber\\
&\quad&-4m^{2}-8m-2\big].
\end{eqnarray}
For systems with large order, i.e., $N_{g}\to\infty$,
\begin{eqnarray}\label{C17}
\langle F\rangle_{g}\sim (N_{g})^{1+\ln2/\ln(m+2)}=
(N_{g})^{2/\tilde{d}}\,.
\end{eqnarray}
This confirms the lower bound for PMFPT provided in for general
graphs~\cite{TeBeVo09}. Thus, for the whole family of the treelike
fractals, in the large $g$ limit, the PMFPT grows as a power-law
function of the network order with the exponent, represented by
$\theta(m)=1+\ln2/\ln(m+2)$, being a decreasing function of $m$.
When $m$ grows from 1 to infinite, the exponent $\theta(m)$ drops
from $1+\ln2/\ln3$ and approaches to 1, indicating that the PMFPT
grows superlinearly with network order. This also means that the
efficiency of the trapping process depends on parameter $m$: the
larger the value of $m$, the more efficient the trapping process.

Before closing this section, we stress that although we only focus
on a special family of fractals, above computational method and
process are also applicable to the trapping problem on other
self-similar trees. For example, we have used this method to compute
the PMFPT to a target hub node in an iterative scale-free tree first
introduced in~\cite{SoHaMa06} and studied in detail
in~\cite{RoHaAv07} and to reproduce the result on PMFPT previously
reported~\cite{ZhXiZhGaGu09}.

\section{Random walks with the trap distributed uniformly on the fractals}

In the Sec. III C, we have studied the PMFPT from a node to a trap
fixed on the central node of treelike fractals averaged over all
possible staring points, and showed that the PMFPT scales
algebraically with network order. In this section, we will
investigate random walks in $T_g$ with the trap uniformly
distributed among all nodes of the fractals. As will be shown, the
EMFPT $\langle H\rangle_{g}$, which is the average of MFPTs over all
pairs of nodes, exhibits the same scaling as that of $\langle F
\rangle_{g}$.

\subsection{Mean first-passage times and effective resistance}

The quantity EMFPT $\langle H \rangle_{g}$ concerned in this case
that trap is randomly selected from all nodes involves a double
average: the first one is over all the starting nodes to a given
trap node and the second one is the average of the first one over
all trap nodes. By definition, $\langle H\rangle_{g}$ is given by
\begin{equation}\label{G1}
\langle
H\rangle_{g}=\frac{1}{N_{g}\left(N_{g}-1\right)}\sum_{i=1}^{N_{g}}\sum_{j=1,j\neq
i}^{N_{g}}F_{ij}(g)\,.
\end{equation}
In general, one can apply the method for pseudoinverse of the
Laplacian matrix~\cite{RaMi71} to calculate numerically but exactly
$\langle H\rangle_{g}$. However, this approach makes heavy demands
on time and computational resources and is thus not tractable for
large networks. To overcome these shortcomings, we will assort to
the theory between electrical networks and random walks to derive
$\langle H \rangle_{g}$, providing a closed-form expression for this
related quantity.



To obtain the solution to $\langle H\rangle_{g}$, we view $T_{g}$ as
resistor networks~\cite{DoSn84} by considering all edges of $T_{g}$
to be unit resistors. Let $R_{ij}(g)$ be the effective resistance
between two nodes $i$ and $j$ in the electrical networks obtained
from $T_{g}$. Then, according to the connection between MFPTs and
effective resistance~\cite{ChRaRuSm89,Te91}, we have
\begin{equation}\label{G2}
F_{ij}(g)+F_{ji}(g)=2\,E_g\,R_{ij}(g)\,,
\end{equation}
using which Eq.~(\ref{G1}) can be recast as
\begin{eqnarray}\label{G4}
\langle
H\rangle_{g}&=&\frac{E_{g}}{N_{g}(N_{g}-1)}\sum_{i=1}^{N_{g}}\sum_{j=1,j\neq
i}^{N_{g}}R_{ij}(g)\nonumber\\
&=&\frac{1}{N_{g}}\sum_{i=1}^{N_{g}}\sum_{j=1,j\neq
i}^{N_{g}}R_{ij}(g),
\end{eqnarray}
where the sum of effective resistance between all pairs of nodes is
the so-called Kirchhoff index~\cite{BoBaLiKl94}, which we denote by
$R_{\rm tot}(g)$. Using the previously obtained
results~\cite{GuMo96,ZhKlLu96}, $R_{\rm tot}(g)$ can be rewritten as
\begin{equation}\label{G5}
R_{\rm tot}(g)=\sum_{i=1}^{N_{g}}\sum_{j=1, j\neq
i}^{N_{g}}R_{ij}(g)=2N_{g}\sum_{i=2}^{N_{g}}\frac{1}{\lambda_{i}(g)},
\end{equation}
where $\lambda_{i}(g)$ $\left(i=2,\dots,N_{g}\right)$ are all the
nonzero eigenvalues of the Laplacian matrix corresponding to
$T_{g}$. Note that Eq.~(\ref{G5}) only holds for a tree. Then, the
EMFPT $\langle H\rangle_{g}$ can be expressed as
\begin{equation}\label{G6}
\langle H\rangle_{g}=2\sum_{i=2}^{N_{g}}\frac{1}{\lambda_{i}(g)}\,.
\end{equation}

We use ${\bf L}_{g}$ to represent the Laplacian matrix for $T_{g}$.
The entries $L_{ij}(g)$ of $\textbf{L}_g$ are defined as follows:
the off-diagonal element $L_{ij}(g)$ is $-1$ if the pair of nodes
$i$ and $j$ are directly linked to each other, otherwise $L_{ij}(g)$
equals $0$, while the diagonal entry $L_{ii}(g)$ is equal to the
degree of node $i$. After reducing the computation of $\langle
H\rangle_{g}$ to determining the sum of the reciprocal of all
nonzero eigenvalues of $\textbf{L}_g$, the next step is to evaluate
this sum.

\subsection{Using eigenvalues of the Laplacian matrix to determine entire mean first-passage time}

Let $P_{g}(\lambda)$ express the characteristic polynomial of matrix
${\bf L}_{g}$, i.e.,
\begin{equation}\label{G7}
P_{g}(\lambda)={\rm det}({\bf L}_{g}-\lambda{\bf I}_{g})\,,
\end{equation}
where ${\bf I}_{g}$ is an $N_g \times N_g$ identity matrix. As
mentioned above, one of the main goals is to find the sum of the
reciprocal of all nonzero eigenvalues of ${\bf L}_{g}$, namely, all
nonzero roots of polynomial $P_{g}(\lambda)$.

In order to determine this sum, we denote ${\bf Q}_{g}$ as a
$(N_g-1) \times (N_g-1)$ sub-matrix of $({\bf L}_{g}-\lambda{\bf
I}_{g})$, which is obtained by removing from $({\bf
L}_{g}-\lambda{\bf I}_{g})$ the row and column corresponding to an
outermost node, e.g., node $K$ in Fig.~\ref{Const2}. Also, we define
${\bf R}_{g}$ as a sub-matrix of $({\bf L}_{g}-\lambda{\bf I}_{g})$
with order $(N_g-2) \times (N_g-2)$, obtained by removing from
$({\bf L}_{g}-\lambda{\bf I}_{g})$ two rows and columns
corresponding to two arbitrary outermost nodes in two different
copies, $T_{g-1}^{(i)}$ ($i=1, 2, \ldots, m+1, m+2$), of $T_{g-1}$
that are constituents of $T_{g}$ (see Fig.~\ref{Const2}). On the
other hand, we denote $Q_{g}(\lambda)$ and $R_{g}(\lambda)$ as the
determinants of ${\bf Q}_{g}$ and ${\bf R}_{g}$, respectively. Then,
the three quantities $P_{g}(\lambda)$, $Q_{g}(\lambda)$, and
$R_{g}(\lambda)$ satisfy the following recursion relations:
\begin{equation}\label{G8}
P_{g+1}(\lambda)=\left|\begin{array}{ccccc}m+2-\lambda & -{\bf
e}_{q} & -{\bf e}_{q} & \cdots & -{\bf e}_{q}
\\-{\bf e}_{q}^{\top} & {\bf Q}_{g} & {\bf O} & \cdots & {\bf O} \\-{\bf e}_{q}^{\top} &
{\bf O} & {\bf Q}_{g} & \cdots & {\bf O} \\\vdots & \vdots & \vdots
& \ & \vdots\\-{\bf e}_{q}^{\top} & {\bf O} & {\bf O} & \cdots &
{\bf Q}_{g}\end{array}\right|,
\end{equation}
\begin{equation}\label{G9}
Q_{g+1}(\lambda)=\left|\begin{array}{ccccc}m+2-\lambda & -{\bf
e}_{r} & -{\bf e}_{q} & \cdots & -{\bf e}_{q}
\\-{\bf e}_{r}^{\top} & {\bf R}_{g}& {\bf O} & \cdots & {\bf O} \\-{\bf e}_{q}^{\top} &
{\bf O} & {\bf Q}_{g} & \cdots & {\bf O} \\\vdots & \vdots & \vdots
& \ & \vdots
\\-{\bf e}_{q}^{\top} & {\bf O} & {\bf O} & \cdots &
{\bf Q}_{g}\end{array}\right|,
\end{equation}
and
\begin{equation}\label{G10}
R_{g+1}(\lambda)=\left|\begin{array}{ccccc}m+2-\lambda & -{\bf
e}_{r} & -{\bf e}_{r} & \cdots & -{\bf e}_{q}
\\-{\bf e}_{r}^{\top} & {\bf R}_{g} & {\bf O} & \cdots & {\bf O} \\-{\bf e}_{r}^{\top} &
{\bf O} & {\bf R}_{g} & \cdots & {\bf O} \\\vdots & \vdots & \vdots
& \ & \vdots
\\-{\bf e}_{q}^{\top} & {\bf O} & {\bf O} & \cdots &
{\bf Q}_{g}\end{array}\right|\,.
\end{equation}
In Eqs.~(\ref{G8})-(\ref{G10}), the superscript $\top$ of a vector
represents transpose and ${\bf e}_{q}$ (${\bf e}_{r}$) is a vector
of order $N_g-1$ ($N_g-2$) with only the first entry being 1 and all
other $N_g-2$ ($N_g-3$) entries equaling 0, i.e.,
\begin{equation}
{\bf e}_{q}=(1,\underbrace{0,0,0,\ldots,0,0}_{N_{g}-2\mbox{}})\,
\end{equation}
and
\begin{equation}
{\bf e}_{r}=(1,\underbrace{0,0,0,\ldots,0,0}_{N_{g}-3\mbox{}})\,.
\end{equation}

In Appendix \ref{AppA}, using the elementary matrix operations we
show that $P_{g+1}(\lambda)$, $Q_{g+1}(\lambda)$, and
$R_{g+1}(\lambda)$ evolve as:
\begin{equation}\label{G11}
P_{g+1}(\lambda)=(m+2)[Q_{g}(\lambda)]^{m+1}
P_{g}(\lambda)+(m+1)\lambda[Q_{g}(\lambda)]^{m+2}\,,
\end{equation}
\begin{eqnarray}\label{G12}
Q_{g+1}(\lambda)&=&[Q_{g}(\lambda)]^{m+2}+(m+1)\lambda
R_{g}(\lambda)[Q_{g}(\lambda)]^{m+1}\nonumber\\
&\quad&+(m+1)R_{g}(\lambda) [Q_{g}(\lambda)]^{m}P_{g}(\lambda)\,
\end{eqnarray}
and
\begin{eqnarray}\label{G13}
R_{g+1}(\lambda)&=&2R_{g}(\lambda)[Q_{g}(\lambda)]^{m+1}+(m+1)\lambda
[R_{g}(\lambda)]^{2}[Q_{g}(\lambda)]^{m}\nonumber\\
&\quad&+m [R_{g}(\lambda)]^{2}[Q_{g}(\lambda)]^{m-1}P_{g}(\lambda).
\end{eqnarray}

After obtaining the recursion relations for $P_{g}(\lambda)$,
$Q_{g}(\lambda)$, and $R_{g}(\lambda)$, next we will compute the sum
of the reciprocal of the non-zero roots of $P_{g}(\lambda)$. Since
$P_{g}(\lambda)$ has one and only one root equal to zero, say
$\lambda_{1}(g)=0$, to find this sum, we define a new polynomial
$P_{g}'(\lambda)$ as
\begin{equation}\label{G14}
P_{g}'(\lambda)=\frac{1}{\lambda}P_{g}(\lambda)\,.
\end{equation}
Obviously, we have
\begin{equation}\label{G15}
\sum_{i=2}^{N_{g}}\frac{1}{\lambda_{i}(g)}=\sum_{i=1}^{N_{g}-1}\frac{1}{\lambda_{i}'(g)},
\end{equation}
where $\lambda_{1}'(g), \lambda_{2}'(g), \ldots,
\lambda_{N_{g}-1}'(g)$ are the $N_{g}-1$ roots of polynomial
$P_{g}'(\lambda)$. Then, we reduce the problem to finding the sum on
the rhs of Eq.~(\ref{G15}).

Note that one can also express the polynomial $P_{g}'(\lambda)$ in
the form of
$P'_{g}(\lambda)=\sum_{j=0}^{N_{g}-1}p'_{g}(j)\lambda^{j}$ [here
$p'_{g}(j)$ is the coefficient of term $\lambda^{j}$ with degree
$j$], such that
\begin{equation}\label{G16}
\sum_{j=0}^{N_{g}-1}p'_{g}(j)\lambda^{j}=p'_{g}(N_{g}-1)\prod_{i=1}^{N_{g}-1}[\lambda-\lambda'_{i}(g)].
\end{equation}
Comparing the coefficients of both sides of Eq.~(\ref{G16}), we have
\begin{equation}\label{G17}
\sum_{i=1}^{N_{g}-1}\frac{1}{\lambda'_{i}(g)}=-\frac{p_{g}'(1)}{p_{g}'(0)}.
\end{equation}
Thus, we turn our aim into determining the two coefficients
$p_{g}'(0)$ and $p_{g}'(1)$.

From Eqs.~(\ref{G11})-(\ref{G13}) we can easily get the recursive
equations,
\begin{equation}\label{G18}
P_{g+1}'(\lambda)=(m+2) [Q_{g}(\lambda)]^{m+1} P_{g}'(\lambda)+(m+1)
[Q_{g}(\lambda)]^{m+2},
\end{equation}
\begin{eqnarray}\label{G19}
Q_{g+1}(\lambda)&=&[Q_{g}(\lambda)]^{m+2}+(m+1)\lambda
R_{g}(\lambda)[Q_{g}(\lambda)]^{m+1}\nonumber\\
&\quad&+(m+1)\lambda
R_{g}(\lambda)[Q_{g}(\lambda)]^{m}P_{g}'(\lambda),
\end{eqnarray}
and
\begin{eqnarray}\label{G20}
R_{g+1}(\lambda)&=&2R_{g}(\lambda)[Q_{g}(\lambda)]^{m+1}+(m+1)\lambda
[R_{g}(\lambda)]^{2}[Q_{g}(\lambda)]^{m}\nonumber\\
&\quad&+m\lambda
[R_{g}(\lambda)]^{2}[Q_{g}(\lambda)]^{m-1}P_{g}'(\lambda).
\end{eqnarray}
Based on these relations we can find the values for $p_{g}'(0)$ and
$p_{g}'(1)$.

We first evaluate $p_{g}'(0)$. For this purpose, let $q_{g}(0)$ and
$r_{g}(0)$ be the constant terms of $Q_{g}(\lambda)$ and
$R_{g}(\lambda)$, respectively. According to
Eqs.~(\ref{G18})-(\ref{G20}), the quantities $p_{g}'(0)$,
$q_{g}(0)$, and $r_{g}(0)$ can be represented recursively as
follows:
\begin{equation}\label{G21}
p_{g+1}'(0)=(m+2)[q_{g}(0)]^{m+1}p'_{g}(0)+(m+1)[q_{g}(0)]^{m+2}\,,
\end{equation}
\begin{equation}\label{G22}
q_{g+1}(0)=[q_{g}(0)]^{m+2}\,,
\end{equation}
and
\begin{equation}\label{G23} r_{g+1}(0)=2r_{g}(0)q_{g}^{m+1}(0).
\end{equation}
Using the initial conditions $p'_{1}(0)=-m-3$, $q_{1}(0)=1$ and
$r_{1}(0)=2$, Eqs.~(\ref{G21})-(\ref{G23}) can be solved to yield
\begin{equation}\label{G24}
p'_{g}(0)=-(m+2)^{g}-1\,,
\end{equation}
\begin{equation}\label{G25}
q_{g}(0)=1\,,
\end{equation}
and
\begin{equation}\label{G26}
r_{g}(0)=2^{g}\,.
\end{equation}

With these obtained results, we go on to compute $p_{g}'(1)$. To
this end, let $q_{g}(1)$ denote the coefficient of term $\lambda$,
i.e., term with degree 1, of polynomial $Q_{g}(\lambda)$. Then,
using Eqs.~(\ref{G18})-(\ref{G20}), we have
\begin{eqnarray}\label{G27}
p_{g+1}'(1)&=&(m+2)
[q_{g}(0)]^{m+1}p'_{g}(1)\nonumber\\
&\quad&+(m+2)(m+1)[q_{g}(0)]^{m}q_{g}(1)p'_{g}(0)\nonumber\\
&\quad&+(m+2)(m+1)[q_{g}(0)]^{m+1}q_{g}(1)
\end{eqnarray}
and
\begin{eqnarray}\label{G28}
q_{g+1}(1)&=&(m+2)
[q_{g}(0)]^{m+1}q_{g}(1)\nonumber\\
&\quad&+(m+1)r_{g}(0)[q_{g}(0)]^{m+1}\nonumber\\
&\quad&+(m+1)r_{g}(0)[q_{g}(0)]^{m}p'_{g}(0).
\end{eqnarray}
Substituting Eqs.~(\ref{G24})-(\ref{G26}) into Eqs.~(\ref{G27}) and
(\ref{G28}), we obtain the explicit expressions for $p_{g}'(1)$ and
$q_{g}(1)$ as
\begin{eqnarray}\label{G29}
p_{g}'(1)&=&\frac{(m+2)^{g-1}}{2m+3}\big[2^{g}(m+2)^{g+2}-2(2^{g}-1)(m+2)^{g+1}\nonumber\\
&\quad&+(2^{g}-1)(m+2)^{g}+(m+2)^{2}-m-2\big],
\end{eqnarray}
\begin{equation}\label{G30}
q_{g}(1)=-(m+2)^{g-1}\left[1-2^{g}+2^{g}(m+2)\right].
\end{equation}

Thus far, we have obtained the intermediate quantities, we now be in
position to determine EMFPT $\langle H\rangle_{g}$. Inserting
Eqs.~(\ref{G24}) and~(\ref{G29}) into Eq.~(\ref{G17}), from which we
can arrive at the explicit formula of the EMFPT $\langle
H\rangle_{g}$ for random walks in $T_g$ as
\begin{eqnarray}\label{G31}
\langle
H\rangle_{g}&=&2\sum_{i=1}^{N_{g}-1}\frac{1}{\lambda'_{i}(g)}=\frac{2(m+2)^{g-1}}{((m+2)^{g}+1)(2m+3)}\nonumber\\
&\quad&\big[2^{g}(m+2)^{g+2}-2(2^{g}-1)(m+2)^{g+1}\nonumber\\
&\quad&+(2^{g}-1)(m+2)^{g}+(m+2)^{2}-m-2\big].
\end{eqnarray}
We have checked this rigorous result given by Eq.~(\ref{G31})
against that obtained via numerical calculations by using the method
of pseudoinverse matrix~\cite{RaMi71}, which fully agree with each
other.

Similar to $\langle F \rangle_{g}$, we can rewrite EMFPT $\langle H
\rangle_{g}$ as a function of network order $N_{g}$ as
\begin{eqnarray}\label{G33}
\langle H\rangle_{g}&=&\frac{2(N_{g}-1)}{(2m^{2}+7m+6)N_{g}}\nonumber\\
&\quad&\big[(m^2+2m+1)(N_{g}-1)^{1+\ln2/\ln(m+2)}\nonumber\\
&\quad&+(2m+3)(N_{g}-1)+m^{2}+3m+2\big].
\end{eqnarray}
Therefore, in the thermodynamic limit ($N_{g}\to\infty$), we have
\begin{equation}\label{G34}
\langle H \rangle_{g} \sim (N_{g})^{1+\ln2/\ln(m+2)}=
(N_{g})^{2/\tilde{d}}\,,
\end{equation}
showing that the EMFPT $\langle H \rangle_{g}$ grows as a power-law
function of the network order $N_g$. It should be mentioned that
Eq.~(\ref{G34}) is in complete agreement with the general result for
fractals given by Eq. (9) in~\cite{TeBeVo09}.


Equations (\ref{C17}) and (\ref{G34}) indicate that $\langle
F\rangle_{g}$ and $\langle H\rangle_{g}$ in the treelike regular
fractals show a similar behavior, both of which grow approximately
as a power-law function of network order $N_{g}$ with the exponent
$\theta(m)= 2/\tilde{d}=1+\ln2/\ln(m+2)$ being a decreasing function
of $m$ and lying between 1 and 2, implying that they both increase
superlinearly with network order. The sameness of scalings for
$\langle F\rangle_{g}$ and $\langle H\rangle_{g}$ in $T_g$ is in
comparison with the results obtained for the deterministic recursive
tree~\cite{ZhQiZhGaGu10}, where the PMFPT to a central node (a node
with highest degree) varies lineally with network order, while the
EMFPT averaged over all node pairs grows with network order $N$ as
$N\ln N$, a scaling larger than the linear one for PMFPT.

Finally, we show the universality of our method for computing EMFPT
in self-similar treelike fractals. Using a process similar to that
above, we determine the EMFPT in the Vicsek fractals, and recover
the exact solution presented in one of our previous
papers~\cite{ZhWjZhZhGuWa10}. 
For details, see Appendix \ref{AppB}.


\section{conclusions}

In this paper, we have presented techniques to determine the
explicit solutions to PMFPT and EMFPT in regular treelike fractals.
First, we provided a formula for the MFPT between two adjacent nodes
in any connected tree, using which one can find PMFPT to a given
target node. We then proposed a method for computing EMFPT, i.e.,
average of MFPTs over all node pairs, which is based on the
relationship of characteristic polynomials of the fractals at
different iterations. It only needs partial knowledge of the
polynomials, but avoids the laborious computation of eigenvalues
that some other methods need. Thus, it is a simple and elegant
method. We have used in detail our techniques to a family of
treelike regular fractals, which include some well-known fractals,
e.g., the $T$ fractal and the Peano basin fractal, as its peculiar
cases. We also applied the methods to other treelike fractals to
demonstrate their universality. 

\begin{acknowledgments}

We would like to thank Xin Li for assistance. This work was
supported by the National Natural Science Foundation of China under
Grant Nos. 60704044 and 61074119, and the Shanghai Leading Academic
Discipline Project No.B114. Y. L also acknowledges the Hui-Chun Chin
and Tsung-Dao Lee Chinese Undergraduate Research Endowment.
\end{acknowledgments}

\appendix

\section{Derivation of recursion relations for relevant polynomials \label{AppA}}

In this appendix, we give the detailed process for the derivation of
recursive relations for polynomials $P_{g}(\lambda)$,
$Q_{g}(\lambda)$, and $R_{g}(\lambda)$. We first derive in detail
the evolution equation for $P_{g}(\lambda)$, while the evolution
relations for $Q_{g}(\lambda)$, and $R_{g}(\lambda)$ can be obtained
analogously.

Note that the vector $(m+2-\lambda, -{\bf e}_{q},-{\bf
e}_{q},\ldots,-{\bf e}_{q})$ of the first row of the matrix on the
rhs of Eq.~(\ref{G8}) can be looked upon as the sum (i.e., a linear
combination with all scalars being 1) of the following $m+3$
vectors: $(m+2-\lambda,{\bf 0}_{q},{\bf 0}_{q},\ldots,{\bf 0}_{q})$,
$(0, -{\bf e}_{q},{\bf 0}_{q},\ldots,{\bf 0}_{q})$, and $(0, {\bf
0}_{q}, -{\bf e}_{q},\ldots,{\bf 0}_{q})$, $\ldots$, $(0, {\bf
0}_{q}, {\bf 0}_{q},\ldots,-{\bf e}_{q})$, where ${\bf 0}_{q}$ is a
zero vector with order $N_g-1$. For all these $m+3$ vectors with
identical order $N_{g+1}$, only one entry is non-zero ($m+2-\lambda$
or $-1$), while the $N_{g+1}-1$ entries are zeros. According to the
properties of determinants, we can rewrite $P_{g+1}(\lambda)$ as a
liner function of the first row on the rhs of Eq.~(\ref{G8}) when
the remaining rows are held fixed. Concretely, $P_{g+1}(\lambda)$
can be expressed as
\begin{widetext}
\begin{eqnarray}\label{G11-1}
P_{g+1}(\lambda)&=&\left|\begin{array}{ccccc}m+2-\lambda & {\bf
0}_{q} & {\bf 0}_{q} & \cdots & {\bf 0}_{q}
\\-{\bf e}_{q}^{\top} & {\bf Q}_{g}& {\bf O} & \cdots & {\bf O} \\-{\bf e}_{q}^{\top} &
{\bf O} & {\bf Q}_{g} & \cdots & {\bf O} \\\vdots & \vdots & \vdots
& \ & \vdots
\\-{\bf e}_{q}^{\top} & {\bf O} & {\bf O} & \cdots &
{\bf Q}_{g}\end{array}\right| + \left|\begin{array}{ccccc}0 & -{\bf
e}_{q} & {\bf 0}_{q} & \cdots & {\bf 0}_{q}
\\-{\bf e}_{q}^{\top} & {\bf Q}_{g} & {\bf O} & \cdots & {\bf O} \\-{\bf e}_{q}^{\top} &
{\bf O} & {\bf Q}_{g} & \cdots & {\bf O} \\\vdots & \vdots & \vdots
& \ & \vdots
\\-{\bf e}_{q}^{\top} & {\bf O} & {\bf O} & \cdots &
{\bf Q}_{g}\end{array}\right|\nonumber\\
&\quad&+\left|\begin{array}{ccccc}0 & {\bf 0}_{q} & -{\bf e}_{q} &
\cdots & {\bf 0}_{q}
\\-{\bf e}_{q}^{\top} & {\bf Q}_{g} & {\bf O} & \cdots & {\bf O} \\-{\bf e}_{q}^{\top} &
{\bf O} & {\bf Q}_{g} & \cdots & {\bf O} \\\vdots & \vdots & \vdots
& \ & \vdots
\\-{\bf e}_{q}^{\top} & {\bf O} & {\bf O} & \cdots &
{\bf Q}_{g}\end{array}\right|+ \dots + \left|\begin{array}{ccccc}0 &
{\bf 0}_{q} & {\bf 0}_{q} & \cdots & -{\bf e}_{q}
\\-{\bf e}_{q}^{\top} & {\bf Q}_{g} & {\bf O} & \cdots & {\bf O} \\-{\bf e}_{q}^{\top} &
{\bf O} & {\bf Q}_{g} & \cdots & {\bf O} \\\vdots & \vdots & \vdots
& \ & \vdots
\\-{\bf e}_{q}^{\top} & {\bf O} & {\bf O} & \cdots &
{\bf Q}_{g}\end{array}\right|.
\end{eqnarray}
\end{widetext}
It is easy to see that the last $m+2$ determinants on the rhs of
Eq.~(\ref{G11-1}) are equal to one another. We use
$P_{g+1}^{(1)}(\lambda)$ to denote the first determinant and use
$P_{g+1}^{(2)}(\lambda)$ to represent any of the last $m+2$
determinants. Thus,
\begin{equation}\label{G11-21}
P_{g+1}(\lambda)=P_{g+1}^{(1)}(\lambda)+(m+2)P_{g+1}^{(2)}(\lambda)\,.
\end{equation}
Below we will calculate these two quantities
$P_{g+1}^{(1)}(\lambda)$ and $P_{g+1}^{(2)}(\lambda)$.

For $P_{g+1}^{(1)}(\lambda)$, it is obvious to have
\begin{equation}\label{G11-2}
P_{g+1}^{(1)}(\lambda)=(m+2-\lambda)[Q_{g}(\lambda)]^{m+2}.
\end{equation}
With regard to $P_{g+1}^{(2)}(\lambda)$, after some transformations
of the matrix corresponding to the determinant, we have
\begin{eqnarray}\label{G11-3}
P_{g+1}^{(2)}(\lambda)&=&[Q_{g}(\lambda)]^{m+1}\left|\begin{array}{cc}0
& -{\bf e}_{q} \\ -{\bf e}_{q}^{\top} &
{\bf Q}_{g}\end{array}\right|\nonumber\\
&=&[Q_{g}(\lambda)]^{m+1}\Bigg (\left|\begin{array}{cc}1-\lambda &
-{\bf e}_{q}
\\ -{\bf e}_{q}^{\top} &
{\bf Q}_{g}\end{array}\right|\nonumber\\
&\quad&+\left|\begin{array}{cc}\lambda-1 & {\bf 0}_{q} \\ -{\bf
e}_{q}^{\top} &
{\bf Q}_{g}\end{array}\right|\Bigg)\nonumber\\
&=&[Q_{g}(\lambda)]^{m+1}[|{\bf
L}_{g}|+(\lambda-1)|{\bf Q}_{g}|]\nonumber\\
&=&[Q_{g}(\lambda)]^{m+1}[P_{g}(\lambda)+(\lambda-1)Q_{g}(\lambda)].
\end{eqnarray}
Plugging Eqs.~(\ref{G11-2}) and~(\ref{G11-3}) into
Eq.~(\ref{G11-21}), we obtain
\begin{eqnarray}\label{G11-4}
P_{g+1}(\lambda)&=&(m+2-\lambda)[Q_{g}(\lambda)]^{m+2}\nonumber\\
&\quad&+(m+2)[Q_{g}(\lambda)]^{m+1}[P_{g}(\lambda)+(\lambda-1)Q_{g}(\lambda)]\nonumber\\
&=&(m+2)[Q_{g}(\lambda)]^{m+1}
P_{g}(\lambda)+(m+1)\lambda[Q_{g}(\lambda)]^{m+2}\,,
\end{eqnarray}
which is exactly Eq.~(\ref{G11}) in the main text.

In a similar way, we can easily derive the recursion relations for
$Q_{g+1}(\lambda)$ and $R_{g+1}(\lambda)$ expressed in
Eqs.~(\ref{G12}) and~(\ref{G13}). The concrete derivation processes
for $Q_{g+1}(\lambda)$ and $R_{g+1}(\lambda)$ are as follows:
\begin{widetext}
\begin{eqnarray}\label{G12-1}
Q_{g+1}(\lambda)&=&[Q_{g}(\lambda)]^{m+1}\left|\begin{array}{cc}m+2-\lambda
& -{\bf e}_{r} \\-{\bf e}_{r}^{\top} & {\bf
R}_{g}\end{array}\right|+(m+1)R_{g}(\lambda)[Q_{g}(\lambda)]^{m}\left|\begin{array}{cc}0
& -{\bf e}_{q} \\-{\bf e}_{q}^{\top} & {\bf Q}_{g}\end{array}\right|\nonumber\\
&=&[Q_{g}(\lambda)]^{m+1}\left [\left|\begin{array}{cc}m+1 & {\bf
0}_{r}
\\-{\bf e}_{r}^{\top} &
{\bf R}_{g}\end{array}\right|+\left|\begin{array}{cc}1-\lambda
& -{\bf e}_{r} \\-{\bf e}_{r}^{\top} & {\bf R}_{g}\end{array}\right|\right]\nonumber\\
&\quad&+(m+1)R_{g}(\lambda)[Q_{g}(\lambda)]^{m}[P_{g}(\lambda)+(\lambda-1)Q_{g}(\lambda)]\nonumber\\
&=&[Q_{g}(\lambda)]^{m+1}\left [(m+1)|{\bf R}_{g}|+|{\bf
Q}_{g}|\right]+(m+1)R_{g}(\lambda)[Q_{g}(\lambda)]^{m}[P_{g}(\lambda)+(\lambda-1)Q_{g}(\lambda)]\nonumber\\
&=&[Q_{g}(\lambda)]^{m+1}[(m+1)R_{g}(\lambda)+Q_{g}(\lambda)]+(m+1)R_{g}(\lambda)[Q_{g}(\lambda)]^{m}[P_{g}(\lambda)+(\lambda-1)
Q_{g}(\lambda)]\nonumber\\
&=&[Q_{g}(\lambda)]^{m+2}+(m+1)\lambda
R_{g}(\lambda)[Q_{g}(\lambda)]^{m+1}+(m+1)R_{g}(\lambda)
[Q_{g}(\lambda)]^{m}P_{g}(\lambda).
\end{eqnarray}
\begin{eqnarray}\label{G13-1}
R_{g+1}(\lambda)&=&[Q_{g}(\lambda)]^{m}\left|\begin{array}{ccc}m+2-\lambda
& -{\bf e}_{r} & -{\bf e}_{r}
\\-{\bf e}_{r}^{\top} & {\bf R}_{g} & {\bf O}
\\ -{\bf e}_{r}^{\top} & {\bf O} & {\bf R}_{g}\end{array}\right|+m[R_{g}(\lambda)]^{2}[Q_{g}(\lambda)]^{m-1}\left|\begin{array}{cc}0 & -{\bf e}_{q} \\-{\bf e}_{q}^{\top} &
{\bf Q}_{g}\end{array}\right|\nonumber\\
&=&[Q_{g}(\lambda)]^{m}\left(\left|\begin{array}{ccc}1-\lambda &
-{\bf e}_{r} & {\bf 0}_{r} \\ -{\bf e}_{r}^{\top} & {\bf R}_{g} & {\bf O} \\
-{\bf e}_{r}^{\top} & {\bf O} & {\bf
R}_{g}\end{array}\right|+\left|\begin{array}{ccc}m+1 & {\bf 0}_{r} & -{\bf e}_{r} \\
-{\bf e}_{r}^{\top} & {\bf R}_{g} & {\bf O}
\\ -{\bf e}_{r}^{\top} & {\bf O} &
{\bf R}_{g}\end{array}\right|\right)\nonumber\\
&\quad&+m
[R_{g}(\lambda)]^{2}[Q_{g}(\lambda)]^{m-1}[P_{g}(\lambda)+(\lambda-1)Q_{g}(\lambda)]\nonumber\\
&=&[Q_{g}(\lambda)]^{m}\left(R_{g}(\lambda)\left|\begin{array}{cc}1-\lambda & -{\bf e}_{r} \\
-{\bf e}_{r}^{\top} & {\bf R}_{g}\end{array}\right|+R_{g}(\lambda)\left|\begin{array}{cc}1-\lambda & -{\bf e}_{r} \\
-{\bf e}_{r}^{\top} & {\bf R}_{g}\end{array}\right|+R_{g}(\lambda)\left|\begin{array}{cc}m+\lambda & {\bf 0}_{r} \\
-{\bf e}_{r}^{\top} & {\bf R}_{g}\end{array}\right|\right)\nonumber\\
&\quad&+m
[R_{g}(\lambda)]^{2}[Q_{g}(\lambda)]^{m-1}[P_{g}(\lambda)+(\lambda-1)Q_{g}(\lambda)]\nonumber\\
&=&[Q_{g}(\lambda)]^{m}\left(2R_{g}(\lambda)|{\bf
Q}_{g}|+(m+\lambda)R_{g}(\lambda)|{\bf R}_{g}|\right)+m
[R_{g}(\lambda)]^{2}[Q_{g}(\lambda)]^{m-1}[P_{g}(\lambda)+(\lambda-1)Q_{g}(\lambda)]\nonumber\\
&=&[Q_{g}(\lambda)]^{m}\{2R_{g}(\lambda)Q_{g}(\lambda) +
(m+\lambda)[R_{g}(\lambda)]^{2}\}+m
[R_{g}(\lambda)]^{2}[Q_{g}(\lambda)]^{m-1}[P_{g}(\lambda)+(\lambda-1)Q_{g}(\lambda)]\nonumber\\
&=&2R_{g}(\lambda)[Q_{g}(\lambda)]^{m+1}+(m+1)\lambda
[R_{g}(\lambda)]^{2}[Q_{g}(\lambda)]^{m}+m
[R_{g}(\lambda)]^{2}[Q_{g}(\lambda)]^{m-1}P_{g}(\lambda)\,,
\end{eqnarray}
\end{widetext}
where ${\bf 0}_{r}$ is a zero vector with order $N_g-2$.

\section{Using recurrence relations for relevant polynomials to determine EMFPT of Vicsek fractals \label{AppB}}

In order to demonstrate the universality of our method for
calculating EMFPT of self-similar tree-like fractals, here we apply
our proposed method to compute the EMFPT for the Vicsek
fractals~\cite{Vi83,BlFeJuKo04}, which is one of the most important
and frequently studied regular fractal classes.

\begin{figure}
\begin{center}
\includegraphics[width=.85\linewidth,trim=100 0 100 0]{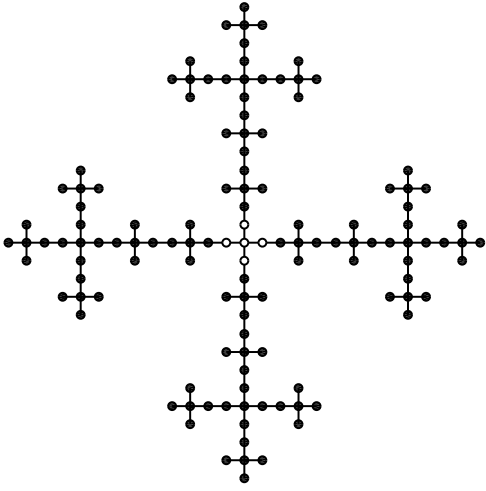}
\caption{The first several iterative processes of a particular
Vicsek fractal $V_{4,3}$. The open circles express the starting
structure $V_{4,1}$.} \label{VicCon}
\end{center}
\end{figure}

The Vicsek fractals are built in an iterative
way~\cite{Vi83,BlFeJuKo04} controlled by the two parameters $f$ and
$g$. Denote by ${\mathbb V}_{f,g}$ $(f\geq 2$, $g\geq1)$ the Vicsek
fractals after $g$ iterations (generations). The initial
construction $(g=1)$ is a starlike cluster composed of $f+1$ nodes
arranged in a crosswise pattern with $f$ peripheral nodes connected
to a central node. This corresponds to ${\mathbb V}_{f,1}$.
Hereafter, we call those nodes farthest from the central node at any
iteration as outmost nodes. For $g\geq 2$, ${\mathbb V}_{f,g}$ is
obtained from ${\mathbb V}_{f,g-1}$. To obtain ${\mathbb V}_{f,g}$,
we generate $f$ identical copies of ${\mathbb V}_{f,g-1}$ (denoted
by ${\mathbb V}_{f,g-1}^{(i)}$, $i=1,2,\ldots, f$) and arrange them
around the periphery of the original ${\mathbb V}_{f,g-1}$ (denoted
by ${\mathbb V}_{f,g-1}^{(0)}$), then we add $f$ new edges, each of
them connects an outmost node in one of the $f$ corner copy
structures and that of the original central structure as shown in
Fig.~\ref{VicCon}.

After introducing the Vicsek fractals, in what follows, we will
apply our method to determine exactly the EMFPT between two nodes
over the whole Vicsek fractal family. Note that in the case without
confusion, we will use the same notations as that used above for the
tree-like fractals $T_{g}$. It should be mentioned that since the
number of recurrence relations for related polynomials increases
with parameter $f$, below we only provide the calculation detail of
the EMFPT for a particular case of $f=3$.

For the convenience of description, we define $V_{g}(\lambda)$ as
the characteristic polynomial of the Laplacian matrix ${\bf L}_{g}$
corresponding to ${\mathbb V}_{3,g}$, i.e.,
\begin{equation}\label{V1}
V_{g}(\lambda)={\rm det}({\bf L}_{g}-\lambda{\bf I}_{g}).
\end{equation}
In order to obtain the sum of the reciprocal of all nonzero
eigenvalues of ${\bf L}_{g}$, viz., all nonzero roots of polynomial
$V_{g}(\lambda)$, we denote $V_{g}^{1}(\lambda)$,
$V_{g}^{2}(\lambda)$, and $V_{g}^{3}(\lambda)$ as determinants of
sub-matrices of $({\bf L}_{g}-\lambda{\bf I}_{g})$, where
$V_{g}^{i}(\lambda)$ ($i=1,2,3$) is obtained by deleting from $({\bf
L}_{g}-\lambda{\bf I}_{g})$ $i$ rows and columns corresponding to
$i$ outmost nodes that belong to $i$ different copy structures
forming ${\mathbb V}_{3,g}$, i.e., ${\mathbb V}_{3,g-1}^{(1)}$,
${\mathbb V}_{3,g-1}^{(2)}$, or ${\mathbb V}_{3,g-1}^{(i)}$.

The polynomials $V_{g}(\lambda)$, $V_{g}^{1}(\lambda)$,
$V_{g}^{2}(\lambda)$, and $V_{g}^{3}(\lambda)$ obey the following
recursive relations:
\begin{eqnarray}\label{V2}
V_{g+1}(\lambda)&=&[V_{g}(\lambda)]^{4}+[6V_{g}^{1}(\lambda)+3V_{g}^{2}(\lambda)+V_{g}^{3}(\lambda)][V_{g}(\lambda)]^{3}\nonumber\\
&\quad&+[9[V_{g}^{1}(\lambda)]^{2}+3V_{g}^{1}(\lambda)V_{g}^{2}(\lambda)][V_{g}(\lambda)]^{2}\nonumber\\
&\quad&+4[V_{g}^{1}(\lambda)]^{3}V_{g}(\lambda)\,,
\end{eqnarray}
\begin{eqnarray}\label{V3}
V_{g+1}^{1}(\lambda)&=&[V_{g}^{1}(\lambda)+V_{g}^{2}(\lambda)][V_{g}(\lambda)]^{3}+[5[V_{g}^{1}(\lambda)]^{2}\nonumber\\
&\quad&+7V_{g}^{1}(\lambda)V_{g}^{2}(\lambda)+[V_{g}^{2}(\lambda)]^{2}+V_{g}^{1}(\lambda)V_{g}^{3}(\lambda)][V_{g}(\lambda)]^{2}\nonumber\\
&\quad&+[5[V_{g}^{1}(\lambda)]^{3}+5[V_{g}^{1}(\lambda)]^{2}V_{g}^{2}(\lambda)]V_{g}(\lambda)\nonumber\\
&\quad&+[V_{g}^{1}(\lambda)]^{4}\,,
\end{eqnarray}
\begin{eqnarray}\label{V4}
V_{g+1}^{2}(\lambda)&=&[V_{g}^{1}(\lambda)+V_{g}^{2}(\lambda)]^{2}[V_{g}(\lambda)]^{2}+[4[V_{g}^{1}(\lambda)]^{3}\nonumber\\
&\quad&+9[V_{g}^{1}(\lambda)]^{2}V_{g}^{2}(\lambda)+4V_{g}^{1}(\lambda)[V_{g}^{2}(\lambda)]^{2}\nonumber\\
&\quad&+[V_{g}^{1}(\lambda)]^{2}V_{g}^{3}(\lambda)]V_{g}(\lambda)+2[V_{g}^{1}(\lambda)]^{4}\nonumber\\
&\quad&+3[V_{g}^{1}(\lambda)]^{3}V_{g}^{2}(\lambda)\,,
\end{eqnarray}
and
\begin{eqnarray}\label{V5}
V_{g+1}^{3}(\lambda)&=&[V_{g}^{1}(\lambda)+V_{g}^{2}(\lambda)]^{3}V_{g}(\lambda)+3[V_{g}^{1}(\lambda)]^{4}\nonumber\\
&\quad&+9[V_{g}^{1}(\lambda)]^{3}V_{g}^{2}(\lambda)+6[V_{g}^{1}(\lambda)]^{2}[V_{g}^{2}(\lambda)]^{2}\nonumber\\
&\quad&+[V_{g}^{1}(\lambda)]^{3}V_{g}^{3}(\lambda).
\end{eqnarray}

Using above recursion relations one can compute the sum of the
reciprocal of the non-zero roots of $V_{g}(\lambda)$. To this end,
we define another polynomial $V_{g}'(\lambda)$ as
\begin{equation}\label{V6}
V_{g}'(\lambda)=\frac{1}{\lambda}V_{g}(\lambda).
\end{equation}
Obviously, the sum of roots of $V_{g}'(\lambda)$ is equal to the sum
of non-zero roots of $V_{g}(\lambda)$. According to the
above-obtained results, we have
\begin{equation}\label{V7}
\langle H \rangle_{g}=-2\frac{v_{g}'(1)}{v_{g}'(0)},
\end{equation}
where $v_{g}'(0)$ is the constant term of $V_{g}'(\lambda)$ and
$v_{g}'(1)$ is the coefficient of term $\lambda$ of
$V_{g}'(\lambda)$. Then, the problem of determining $\langle H
\rangle_{g}$ is reduced to finding the quantities $v_{g}'(0)$ and
$v_{g}'(1)$.

From Eqs.~(\ref{V2})-(\ref{V5}) we can easily get the following four
recursive equations:
\begin{eqnarray}\label{V8}
V_{g+1}'(\lambda)&=&\lambda^{3}[V_{g}'(\lambda)]^{4}+\lambda^{2}[6V_{g}^{1}(\lambda)+3V_{g}^{2}(\lambda)+V_{g}^{3}(\lambda)]\nonumber\\
&\quad&[V_{g}'(\lambda)]^{3}+\lambda[9[V_{g}^{1}(\lambda)]^{2}+3V_{g}^{1}(\lambda)V_{g}^{2}(\lambda)][V_{g}'(\lambda)]^{2}\nonumber\\
&\quad&+4[V_{g}^{1}(\lambda)]^{3}V_{g}'(\lambda)\,,
\end{eqnarray}
\begin{eqnarray}\label{V9}
V_{g+1}^{1}(\lambda)&=&\lambda^{3}[V_{g}^{1}(\lambda)+V_{g}^{2}(\lambda)][V_{g}'(\lambda)]^{3}+\lambda^{2}[5[V_{g}^{1}(\lambda)]^{2}\nonumber\\
&\quad&+7V_{g}^{1}(\lambda)V_{g}^{2}(\lambda)+[V_{g}^{2}(\lambda)]^{2}+V_{g}^{1}(\lambda)V_{g}^{3}(\lambda)][V_{g}'(\lambda)]^{2}\nonumber\\
&\quad&+\lambda[5[V_{g}^{1}(\lambda)]^{3}+5[V_{g}^{1}(\lambda)]^{2}V_{g}^{2}(\lambda)]V_{g}'(\lambda)\nonumber\\
&\quad&+[V_{g}^{1}(\lambda)]^{4}\,,
\end{eqnarray}
\begin{eqnarray}\label{V10}
V_{g+1}^{2}(\lambda)&=&\lambda^{2}[V_{g}^{1}(\lambda)+V_{g}^{2}(\lambda)]^{2}[V_{g}'(\lambda)]^{2}+\lambda[4[V_{g}^{1}(\lambda)]^{3}\nonumber\\
&\quad&+9[V_{g}^{1}(\lambda)]^{2}V_{g}^{2}(\lambda)+4V_{g}^{1}(\lambda)[V_{g}^{2}(\lambda)]^{2}\nonumber\\
&\quad&+[V_{g}^{1}(\lambda)]^{2}V_{g}^{3}(\lambda)]V_{g}'(\lambda)+2[V_{g}^{1}(\lambda)]^{4}\nonumber\\
&\quad&+3[V_{g}^{1}(\lambda)]^{3}V_{g}^{2}(\lambda)\,,
\end{eqnarray}
and
\begin{eqnarray}\label{V11}
V_{g+1}^{3}(\lambda)&=&\lambda[V_{g}^{1}(\lambda)+V_{g}^{2}(\lambda)]^{3}V_{g}'(\lambda)+3[V_{g}^{1}(\lambda)]^{4}\nonumber\\
&\quad&+9[V_{g}^{1}(\lambda)]^{3}V_{g}^{2}(\lambda)+6[V_{g}^{1}(\lambda)]^{2}[V_{g}^{2}(\lambda)]^{2}\nonumber\\
&\quad&+[V_{g}^{1}(\lambda)]^{3}V_{g}^{3}(\lambda)\,.
\end{eqnarray}
Based on these relations we can find the values for $v_{g}'(0)$ and
$v_{g}'(1)$.

Let us first evaluate $v_{g}'(0)$. To this end, we use
$v_{g}^{1}(0)$, $v_{g}^{2}(0)$, and $v_{g}^{3}(0)$ to denote
separately the constant terms of $V_{g}^{1}(\lambda)$,
$V_{g}^{2}(\lambda)$ and $V_{g}^{3}(\lambda)$. Making use of
Eqs.~(\ref{V8})-(\ref{V11}), we can derive the following important
relations:
\begin{equation}\label{V12}
v_{g+1}'(0)=4[v_{g}^{1}(0)]^{3}v_{g}'(0)\,,
\end{equation}
\begin{equation}\label{V13}
v_{g+1}^{1}(0)=[v_{g}^{1}(0)]^{4}\,,
\end{equation}
\begin{equation}\label{V14}
v_{g+1}^{2}(0)=2[v_{g}^{1}(0)]^{4}+3[v_{g}^{1}(0)]^{3}v_{g}^{2}(0)\,,
\end{equation}
and
\begin{eqnarray}\label{V15}
v_{g+1}^{3}(0)&=&3[v_{g}^{1}(0)]^{4}+9[v_{g}^{1}(0)]^{3}v_{g}^{2}(0)\nonumber\\
&\quad&+6[v_{g}^{1}(0)]^{2}[v_{g}^{2}(0)]^{2}+[v_{g}^{1}(0)]^{3}v_{g}^{3}(0)\,.
\end{eqnarray}
Considering $v_{1}'(0)=-4$, $v_{1}^{1}(0)=1$, $v_{1}^{2}(0)=2$, and
$v_{1}^{3}(0)=3$, Eqs.~(\ref{V12})-(\ref{V15}) are resolved to
obtain
\begin{equation}\label{V16}
v_{g}'(0)=-4^{g}\,,
\end{equation}
\begin{equation}\label{V17}
v_{g}^{1}(0)=1\,,
\end{equation}
\begin{equation}\label{V18}
v_{g}^{2}(0)=3^{g}-1\,,
\end{equation}
and
\begin{equation}\label{V19}
v_{g}^{3}(0)=\frac{3}{4}(3^{g}-1)^{2}\,.
\end{equation}

With these obtained results, we continue to calculate $v_{g}'(1)$.
To achieve this goal, we define $v_{g}^{1}(1)$ as the coefficient of
term $\lambda$ of polynomial $V_{g}^{1}(\lambda)$. According to
Eqs.~(\ref{V8})-(\ref{V11}), the quantities $v_{g}'(1)$ and
$v_{g}^{1}(1)$ can be represented recursively as follows:
\begin{eqnarray}\label{V20}
v_{g+1}'(1)&=&4[v_{g}^{1}(0)]^{3}v_{g}'(1)+12[v_{g}^{1}(0)]^{2}v_{g}'(0)v_{g}^{1}(1)\nonumber\\
&\quad&+9[v_{g}^{1}(0)]^{2}[v_{g}'(0)]^{2}+3v_{g}^{1}(0)v_{g}^{2}(0)[v_{g}'(0)]^{2}
\end{eqnarray}
and
\begin{eqnarray}\label{V21}
v_{g+1}^{1}(1)&=&4[v_{g}^{1}(0)]^{3}v_{g}^{1}(1)+5[v_{g}^{1}(0)]^{3}v_{g}'(0)\nonumber\\
&\quad&+5[v_{g}^{1}(0)]^{2}v_{g}^{2}(0)v_{g}'(0).
\end{eqnarray}
Plugging Eqs.~(\ref{V16})-(\ref{V18}) into Eqs.(\ref{V20}) and
(\ref{V21}) and using the initial conditions $v_{1}'(1)=9$ and
$v_{1}^{1}(1)=-5$, we obtain the exact solutions to $v_{g}'(1)$ and
$v_{g}^{1}(1)$,
\begin{equation}\label{V22}
v_{g}'(1)=\frac{2^{2g-3}}{11}(21\times12^{g}-11\times4^{g}-10)
\end{equation}
and
\begin{equation}\label{V23}
v_{g}^{1}(1)=-5\times2^{2g-3}(3^{g}-1)\,.
\end{equation}

After obtaining all the intermediate quantities, we can determine
the EMFPT of the fractal ${\mathbb V}_{3,g}$. Inserting Eqs.
(\ref{V16}) and (\ref{V22}) into Eq.(\ref{V7}), we obtain the
explicit formula for the EMFPT $\langle H \rangle_{g}$ in ${\mathbb
V}_{3,g}$ as
\begin{equation}\label{V24}
\langle
H\rangle_{g}=\frac{1}{44}(21\times12^{g}-11\times4^{g}-10)\,,
\end{equation}
which is completely consistent with the result previously reported
in~\cite{ZhWjZhZhGuWa10}.

\nocite{*}


\end{document}